\documentclass[sigconf,screen]{acmart}

\AtBeginDocument{%
  \providecommand\BibTeX{{%
    Bib\TeX}}}
\setcopyright{rightsretained}
\acmDOI{10.1145/3664646.3664762}
\acmYear{2024}
\copyrightyear{2024}
\acmSubmissionID{fsews24aiwaremain-p28-p}
\acmISBN{979-8-4007-0685-1/24/07}
\acmConference[AIware '24]{Proceedings of the 1st ACM International Conference on AI-Powered Software}{July 15--16, 2024}{Porto de Galinhas, Brazil}
\acmBooktitle{Proceedings of the 1st ACM International Conference on AI-Powered Software (AIware '24), July 15--16, 2024, Porto de Galinhas, Brazil}
\received{2024-04-05}
\received[accepted]{2024-05-04}

\usepackage{cite}
\usepackage{algorithmic}
\usepackage{graphicx}
\usepackage{tabularx}
\usepackage{url}
\usepackage{booktabs}
\usepackage{textcomp}
\usepackage{xcolor}
\def\BibTeX{{\rm B\kern-.05em{\sc i\kern-.025em b}\kern-.08emT\kern-.1667em\lower.7ex\hbox{E}\kern-.125emX}}

\usepackage[framemethod=TikZ]{mdframed}
\definecolor{backgroundpage}{HTML}{F8F5F4}

\definecolor{prompt}{HTML}{BCBCF2}

\definecolor{chatgpt4}{HTML}{ffb6c1}
\definecolor{chatgpt3}{HTML}{b0e0e6}
\definecolor{bard}{HTML}{daa520}
\definecolor{copilot}{HTML}{708090}

\newenvironment{coloredframe}[2][]{
\mdfsetup{
skipabove=3pt, skipbelow=3pt,% pads top.
innerlinewidth=1.5pt, 
innerlinecolor=white, % size of bar
linewidth=0pt,
backgroundcolor=#2!15,
innerleftmargin=10pt, % Padding on the left
innerrightmargin=15pt,      % Padding on the right
innertopmargin=10pt,        % Padding at the top
innerbottommargin=10pt,     % 
roundcorner=5pt           % Radius of the round 
}
\begin{mdframed}}
{\end{mdframed}}

\newenvironment{zeroindent}
  {\par\setlength{\parindent}{0pt}}
  {\par}

\newcommand{\dialoguegpt}[2]{    
    \begin{coloredframe}{#1}
    \begin{zeroindent} #2 \end{zeroindent}
    \end{coloredframe}
}


\begin{document}

\title{Unveiling Assumptions: Exploring the Decisions of AI Chatbots and Human Testers}

\author{Francisco Gomes de Oliveira Neto}
\orcid{0000-0001-9226-5417}
\affiliation{%
  \institution{Chalmers - University of Gothenburg}
  \city{Gothenburg}
  \country{Sweden}
}
\email{francisco.gomes@cse.gu.se}

\begin{abstract}
The integration of Large Language Models (LLMs) and chatbots introduces new challenges and opportunities for decision-making in software testing. Decision-making relies on a variety of information, including code, requirements specifications, and other software artifacts that are often unclear or exist solely in the developer's mind. To fill in the gaps left by unclear information, we often rely on assumptions, intuition, or previous experiences to make decisions. This paper explores the potential of LLM-based chatbots like Bard, Copilot, and ChatGPT, to support software testers in test decisions such as prioritizing test cases effectively. We investigate whether LLM-based chatbots and human testers share similar ``assumptions'' or intuition in prohibitive testing scenarios where exhaustive execution of test cases is often impractical. Preliminary results from a survey of 127 testers indicate a preference for diverse test scenarios, with a significant majority (96\%) favoring dissimilar test sets. Interestingly, two out of four chatbots mirrored this preference, aligning with human intuition, while the others opted for similar test scenarios, chosen by only 3.9\% of testers. Our initial insights suggest a promising avenue within the context of enhancing the collaborative dynamics between testers and chatbots.
\end{abstract}


\maketitle

\section{Introduction}

% What roles do autonomous agents play in the development and maintenance of software in the FM era?xx

% How do we debug and monitor AIware in the FM era?

% How should we change SE curriculum, training and mentoring in the FM era?

% How does AIware impact developer productivity?

%I would have presented a case in which a non-obvious choice has to be made, with a clear tradeoff.

% NOte that there is not a correct option.

Artificial intelligence (AI) offers promising new avenues for enhancing decision-making processes in software testing, particularly through the use of generative AI (GAI) technologies. These tools have the potential to transform how testers work with test case design~\citep{nashid2023retrieval}, program repair~\citep{sobania2023bugrepair}, or debugging~\citep{kang2023large}, particularly when using Agentware by, for instance, combining Large Language Models (LLM) and chatbots (Agentware). The conversational aspect between a human and a potential Testing Agent opens up the possibilities for collaborations that can enhance the testing competence of both the human and the chatbot~\citep{feldt2023autonomous}, thus, challenging our knowledge of roles that Agentware play in the testing and evolution of software in this new era software craftsmanship.

With LLM-based chatbots\footnote{For brevity, we use the term chatbots and LLM-based chatbots interchageably in the remainder of this paper.} such as Bard\slash Gemini, Copilot, and ChatGPT, there is a unique opportunity to leverage these technologies in supporting software testers in making more informed decisions about which test cases to execute. For instance, in regression testing where the exhaustive execution of all test cases is infeasible due to resource constraints, testers often need to select or prioritise subsets of test cases~\citep{yoo2012survey,oliveiraNeto2018apsec}. Similarly, designing tests for all features is impractical, thus testers frequently need to determine which functionalities to test based on the assumption that common issues found in previous versions or similar systems are likely to reoccur. Those choices can be done manually (based on a testers' experience or ``gut feeling''), automatically by an algorithm, or a combination of both.

\begin{coloredframe}{gray}
\textbf{Our vision is that:} Agentware will allow us to explore and harness the ``human'' oracles to enhance the experience of testers (artificial \textit{and} human) via a feedback cycle of questions and corresponding answers\slash discoveries triggered by, e.g., prompt engineering, hallucinations, or other types of interactions between both parties.    
\end{coloredframe}

Our initial investigation explores a simple yet common scenario in test prioritization. We aim to demonstrate how Agentware can collaborate with humans by recommending test subsets to support the decision-making process in prioritizing and selecting tests. By exploring the capabilities of chatbots in aiding tests in this context, this paper offers insights on whether chatbots and human testers often share similar intuitions about test selection. Therefore, we discuss the preliminary results of a study where we investigate the question: \textit{Do testers and LLM-based chatbots rely on similar assumptions\slash intuition when prioritising tests?}

Our preliminary results reveal that testers consistently choose dissimilar tests scenarios, and favour those subsets with more different tests (96\% of our 127 respondents). When presenting the same problem to four chatbots, two of them (Copilot and ChatGPT 4.0) answered similar to our sample of testers, whereas the remaining chatbots chose the subset with similar test scenarios (ChatGPT3.5 and Bard\footnote{The data was collected before Google renamed Bard to Gemini, therefore, we refer to Bard throughout the paper.}) which was only chosen by 3.9\% of the testers. Nonetheless, a qualitative assessment of the justifications provided by both chatbots and humans raised similar reasoning behind their respective choices, such as ensuring coverage of the scenarios they assumed would be more executed by an end-user, test cases that would help the tester become familiar with the System Under Test (SUT), or favouring dissimilar scenarios. Interestingly, neither of those points were inluded in the prompt or description given to our participants (artificial or humans). Contrastingly, one of the test scenarios involved the creation of a random hero\footnote{In the provided SUT, a random hero \textit{could be seen} as a hero character for a game with randomised properties (e.g., class, race, attributes).} which was only raised as risky scenario by the human testers.

We discuss our preliminary findings in relation to the concept of Testing Agents and the SocraTest framework introduced by Feldt et al.\ which advocates for the collaboration between testers and chatbots to enhance their respective domain and testing expertise~\citep{feldt2023autonomous}. 

%Therefore, we outline the relationship between our initial findings and existing literature on LLM-based chatbots supporting software engineering tasks (Section~\ref{sec:background}), followed by a description of our survey of testers (Section~\ref{sec:survey}) and its comparison with the results obtained from chatbots (Section~\ref{sec:chatbots}). Lastly, we summarise our vision and outline future work in this area (Section~\ref{sec:conclusions}).

\section{Software Testing and AI Chatbots}
\label{sec:background}

Large Language Models (LLMs) based chatbots, such as ChatGPT, have significantly impacted software engineering, particularly in generating code, test scripts, and providing insights into software engineering concepts, techniques, and tools. Chatbots are increasingly being applied to software testing tasks, evolving into conversational agents that assist testers in decision-making processes~\citep{santos2023testing,wang2024software,kang2023large,sobania2023bugrepair,feldt2023autonomous}, thereby presenting new research opportunities in mitigating risks associated with these AI-driven interventions~\citep{sallou2023breaking}.

Nonetheless, Agentware introduce unique challenges such as response variability, ethical concerns, intellectual property, and biased outputs~\citep{sallou2023breaking,ahmad2023collaborative}. Sarkar et al. report on some of these challenges in a user study which non-expert end user programmers use LLM-assisted tools for solving data tasks in spreadsheets~\citep{sarkar2022like}. When describing the collaboration between LLM-based chatbots and humans, authors draw an interesting parallel to the roles of ``driver'' and ``navigator'' in pair programming. Particularly, authors describe that a solo programmer working with an AI assistant could seamlessly shift roles between leading the task and guiding the process, moment by moment~\citep{sarkar2022like}. This fluidity suggests that when employed as Testing Agents, chatbots may facilitate a similar range of interaction dynamics.

In fact, Feldt et al. propose a taxonomy to characterize those Testing Agents based on their autonomy levels, and aid the testers by, e.g., recommending surprising test input, or even make them question their understanding of the SUT's specification based on potential hallucinations of the LLM~\citep{feldt2023autonomous}. Our initial study leverages on the ``Contextual Prompting'' level from Feldt et al.'s taxonomy to explore the interaction between human testers and LLMs without direct dialogue. Our preliminary findings suggest that, despite differences in specific recommendations, the rationale provided by chatbots often mirrors the intuition of human testers, particularly, the importance of test diversity.

This alignment indicates potential for greater synergy at higher autonomy levels within Feldt et al.'s taxonomy, hinting at the intriguing possibility that human testers and LLMs could share similar testing strategies and priorities. In fact, future research that replicate our study with more advanced levels of their taxonomy (e.g., at the level of Conversational Testing with Tools) could uncover valuable insights into the instincts of both humans and LLMs regarding their recommendations on software testing tasks.

\section{Survey Insights: Testers' Strategies for Choosing Test Cases}
\label{sec:survey}

Literature has shown the benefits of using diversity as a criteria for generating~\citep{feldt2008diversity,henard2016icse}, and prioritising test cases~\citep{oliveiraNeto2018ast}. Results from those studies indicate an underlying intuition that selecting diverse test cases typically lead to more cost-effective testing. We began by investigating this assumption with humans, before comparing it with AI chatbots.

During live presentations delivered at various companies, we conducted surveys among the attendees to gather insights on their preferences in a simple yet typical test selection problem when confronted with limited resources for test execution. To facilitate anonymous data collection, we employed Mentimeter\footnote{\url{https://mentimeter.com/}} during these presentations, ensuring that participants were informed about the data collection process and given the option to opt out at any stage. Out of 135 respondents, 127 consented to participate in our study, providing a substantial dataset for analysis. We collected data from a total of 127 software engineers working in quality assurance in two companies in Europe, across three separate sessions --- two in June 2019 (Company A, 110 respondents) and one in April 2022 (Company B, 17 respondents).

We compare whether there is a similar ``intuition'' between our sample of respondents and the LLM-based chatbots in their recommendations for testing executions. Therefore, we prompt four different LLM-based chatbots with the same problems shared to practitioners in our past presentations and compare: (i) their choices of test cases, and (ii) the reasoning behind those choices. The following selection problem was described to the testers and the chatbots.\\

\dialoguegpt{prompt}{
\textbf{Description of the scenario:} Imagine that you are Natalie, a software tester that started at a company today. You are responsible for executing manual test cases for a game. You started today, so you don't know much about the system yet. You were asked to execute the following three test cases:\\

\textbf{TC1: New game (new hero)} \\
Step 1: Select "New game" \\
Step 2: Create a new hero \\
Step 3: Save hero \\
Step 4: Start game \textbf{End of TC1} \\
\\
\textbf{TC2: New game (random hero)} \\
Step 1: Select "New game" \\
Step 2: Generate a random hero \\
Step 3: Save hero \\
Step 4: Start game \textbf{End of TC2} \\
\\
\textbf{TC3: Load game} \\
Step 1: Select "Load game" \\
Step 2: Select a previous game file \\
Step 3: Load the game \textbf{End of TC3} \\

Natalie realises that there is little time left until she needs to leave work for the weekend and that she will only be able to execute two out of the three test cases above. If you were Natalie, which pair of tests would you choose? Explain why did you choose that pair.\\

\textbf{Choice A:} TC1 and TC2 \\
\textbf{Choice B:} TC1 and TC3 \\
\textbf{Choice C:} TC2 and TC3 
}

Our goal was to find initial evidence of a pattern between humans and the artificial agents to, then, expand that investigation in a controlled experiment. The goal of this study is \textit{not} to investigate the impact of different ways to prompt these models and their recommendations (e.g., changing the order of the options or including reasoning to the LLM), nor is it to explore the scalability of the scenario by providing more variations to the problem (e.g., more test cases, larger subsets, different versions of similar SUTs). Additionally, since the human testers' data was collected before these models were released, we aimed to keep the prompt as close as possible to the description given to participants.

Despite our efforts to ensure consistency in the execution of our study, working with LLMs introduce a series of challenges for the reliability of the recommendations produced. Below, we list the limitations in our data collection, particularly regarding the LLM interaction. We used the recommendations from Sallou et al.~\citep{sallou2023breaking} to mitigated those limitations in our analysis.\\

\textbf{Limited reproducibility:} We explore the chat aspect of the LLM, such that parameters like temperature are not directly configured by an end-user, which leads to output variability and time-based output drift. We mitigate those threats by (i) assessing output variability by (manually) prompting the model five times, (ii) providing execution meta-data (dates, prompts and corresponding answers  in our data package~\citep{replication2024aist}), and (iii) prompting the model on the same day with the exact same prompt, but resetting the sessions and chatbot at each repetition.\\

\textbf{Simplicity of the example:} The problem presented is simplified to avoid cognitive overload for  our human respondents during those presentations. On the other hand, we wanted to be fair with the LLM by not introducing variations in the example (e.g., adding more information or changing the order of options) to minimise variations from the data collected with humans.\\

\begin{figure}[H]
    \centering
    \includegraphics[width=\linewidth]{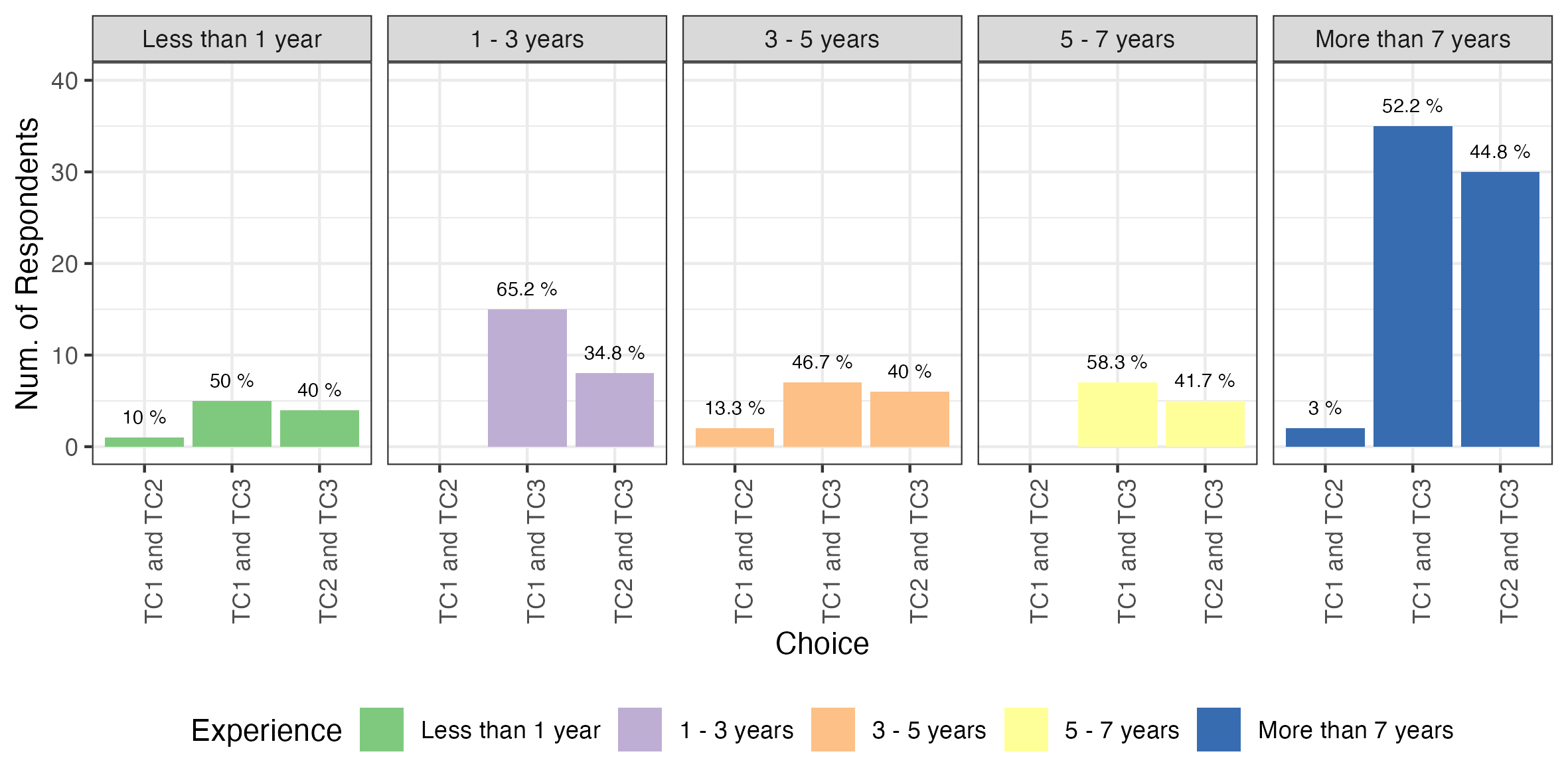}
    \caption{Survey results per experience of the respondents. For each level, the last year is not inclusive --- e.g., participants had between one year (inclusive) and 3 years (exclusive).}
    \label{fig:survey}
\end{figure}

Note that \textit{there is not} an actual\slash expected answer for Natalie's problem because ``choosing the best'' subset of tests depends on domain information that we \textit{deliberately} left out to trigger testers to apply their ``gut feeling'' and fill in those gaps with their intuition (e.g., testing goals, importance of different requirements, personal experience, common usage scenarios). Figure~\ref{fig:survey} shows the participants' answers and a clear preference, among human testers, for selecting dissimilar types of tests. Specifically, 53.5\% of respondents chose a combination of TC1 and TC3, while 42.6\% preferred TC2 and TC3, and only a minimal 3.9\% opted for TC1 and TC2. This pattern persisted across testers of all experience levels, indicating a widespread recognition of the value of test diversity, as reported in literature.

The rationale behind these choices was particularly insightful; testers who favored TC2 over TC1 often cited the importance of testing randomness due to its potential for uncovering bugs or believed that TC2 offered similar coverage to TC1 but with added benefits due to its randomness. As mentioned, the survey intentionally excluded specific information about the System Under Test (SUT), such as coverage or requirements, to prevent bias and encourage testers to rely on their judgment. This approach also reflects a simple yet realistic scenario where testers start working in projects with limited understanding of the SUT, using test scripts like those in our survey to form a clearer picture of the SUT.

\section{Comparative Analysis of Chatbots}
\label{sec:chatbots}

Below, we report on the responses of various generative AI chatbots, including ChatGPT versions 3.5 and 4, Bard, and Copilot\footnote{We prompted Copilot in all of its three UI configurations: More Creative, More Balanced, and More Precise. There were \textit{no} notable differences between all three configurations in their chosen pair of tests, or in the content. Therefore, for brevity, we discuss only the results of the More Balanced configuration.}. We prompted each chatbot with the same description presented to practitioners. Each chatbot was prompted five times to observe variance in responses, which remained consistent across iterations with small variations in the chatbots' reasoning. For brevity, we summarise the content and reasoning for each chatbot followed by a compressed version of their response. We make the complete data with (i) our survey, (ii) participants' answers, and (iii) the chatbots' answers in a Zenodo package for this study~\citep{replication2024aist}.\footnote{\url{https://zenodo.org/records/10622332}}\\

\textbf{ChatGPT 4.0's Choices:} ChatGPT 4.0 consistently selected TC1 and TC3 across all trials, citing the coverage of major functionalities, diversity of test scenarios, and strategic prioritization as its reasons. This selection was justified by the potential impact on player retention and satisfaction, specifically mentioning that TC2 was redundant due to ``\textit{the core functionality tested here --- starting a new game and saving the hero --- is already covered by TC1}''. These justifications were in line with those provided by human testers who made similar choices.\\

\dialoguegpt{chatgpt4}{
\textbf{ChatGPT 4.0}:

If I were Natalie, I would choose \textbf{Option 2: TC1 and TC3.} Here's why: \\

\textbf{1. Coverage of Different Functional Areas:} TC1 and TC3 cover two distinct functional areas of the game. [...] TC1 tests the functionality of starting a new game with a completely new character [...]  [TC3 is crucial for assessing the persistence and retrieval aspects of the game's data management.] ...\\

\textbf{2. Diversity in Test Scenarios:} [...] While TC1 and TC2 are quite similar [...], adding TC3 introduces a completely different scenario that involves game continuity and data integrity. This diversity in testing scenarios can help uncover a wider range of issues related to both game initialization and continuation.\\

\textbf{3. Efficient Use of Time:} [...] Executing TC1 and TC2, which are quite similar, might not be as time-efficient in terms of uncovering unique bugs or issues within the system.\\

\textbf{4. Strategic Importance for Early Testing:} [...] These functionalities are likely to be used frequently by end-users, [...] critical for a good user experience. [...]
}

\textbf{ChatGPT 3.5's Choices:} Conversely, ChatGPT 3.5 primarily recommended executing TC1 and TC2, with one instance of suggesting TC1 and TC3 instead. Its rationale emphasized that ``\textit{new game creation is one of the fundamental aspects of the game}'', while loading the game (TC3) `\textit{`can be considered secondary to the critical functionality of starting a new game, especially when time is limited}''. ChatGPT 3.5 also justified test diversity as one of the reasons for choosing TC1 and TC2, by using ``randomness'' as the distinguishing aspect between that pair. Ironically, diversity of test scenarios was \textit{also} its main argument in the single instance in which ChatGPT 3.5 recommended TC1 and TC3.

\dialoguegpt{chatgpt3}{
\textbf{ChatGPT 3.5}:

If I were Natalie, I would choose \textbf{Option 1: TC1 and TC2.} Here's why:\\

\textbf{1. Coverage:} By selecting TC1 and TC2, you cover both scenarios related to starting a new game, which is a fundamental functionality in the game. [...]\\

\textbf{2. Priority:} Starting a new game is typically a higher priority for testing because it's a common and user action. [...]\\

\textbf{3. Time constraints:} Since Natalie has limited time before the weekend and can only execute two out of the three test cases, focusing on the two test cases related to starting a new game is a more practical choice. [...]\\

While TC3 (Load game) is also important, it can be considered a secondary functionality compared to starting a new game, and it may be less likely to have critical issues that block players from enjoying the game initially.[...]
}

\textbf{Bard's Recommendations} Bard consistently recommended TC1 and TC2 in all trials, suggesting this combination would ``\textit{maximize [Natalie's] information gain within the limited time, covers core new game functionality with variations, and optimizes her testing path for efficiency.}'' Bard's rationale similarly downplayed the importance of loading a game (TC3), suggesting it offered less exposure to core gameplay scenarios due to the lack of saved files to execute TC3.\\

\dialoguegpt{bard}{
\textbf{Bard}: 
If I were Natalie, I would choose \textbf{Option 1: TC1 and TC2}. Here's why:\\

1) Covers new game functionality: [...]; 2) Covers different hero creation methods: [...]; 3) Provides more data on hero creation flows: Since Natalie started today, [the core gameplay loop, ..., is valuable for familiarization.]; 4) Both are independent test cases: Running TC1 and TC2 doesn't depend on the availability of a previous save file, making them more efficient to execute with limited time. \\

\textbf{Considerations against other options:}

\textbf{Option 2 (TC1 \& TC3):} While TC3 tests loading functionality, it relies on having a previously saved game[...]

\textbf{Option 3 (TC2 \& TC3):} [...] both TC2 and TC3 focus on hero creation (one random, one manual), offering less diverse testing compared to Option 1.\\

[...] It's important to remember that this is just one possible approach, and depending on the specific context and information about the game and test cases, another option might be better suited.
}

\textbf{Copilot's Choices:} Across all configurations, Copilot showed no significant variance in its responses, uniformly advising the execution of TC1 and TC3. It reasoned that this choice ``\textit{ensures that fundamental features are thoroughly tested before the weekend begins}''. Copilot mentions that ``\textit{while TC2 (New game with a random hero) is interesting, it doesn't directly impact the core gameplay flow}''. Unlike other chatbots, Copilot's rationale did not explicitly reference test diversity, risk management, or time efficiency.

\dialoguegpt{copilot}{
\textbf{Copilot (More Balanced) - Trial 1}: 

As Natalie, I would choose \textbf{Option 2: TC1 and TC3}. Here's why:

1) TC1 ensures that the basic functionality works correctly.

2) TC3 is crucial for players who want to continue their progress. ...\\

While TC2 (New game with a random hero) is interesting, it doesn't directly impact the core gameplay flow.[...]
}

\textbf{Chatbot Comparisons:} In summary, ChatGPT 4.0 and Copilot aligned with the majority of respondents in preferring TC1 and TC3, highlighting their emphasis on fundamental feature testing. In contrast, Bard and ChatGPT 3.5's preference for TC1 and TC2 was less chosen by human testers, though their reasoning matched practitioner insights on prioritizing core gameplay functionalities. Surprisingly, none of the chatbots in \textit{any} of our 30 trials opted for TC2 and TC3, a choice made by a significant portion of testers (42.6\%) based on concerns over the potential bugs in hero creation randomness compared to TC1.

\section{Conclusions and Future Work}
\label{sec:conclusions}

This paper shares preliminary findings on employing Agentware to assist testers in making decisions. We introduced a test prioritization problem to 127 human testers with diverse experience levels and four LLM-based chatbots, comparing their answers. Key highlights of our study are:

\begin{itemize}
    \item Testers of all experience levels and two chatbots (Copilot and ChatGPT 4.0) preferred diverse test scenarios, aligning with literature on the effectiveness of varied test suites for bug detection.
    \item Conversely, ChatGPT 3.5 and Bard focused on similar scenarios, a choice echoed by a minority of human testers (3.9\%).
    \item Despite showing variability in responses, especially ChatGPT 3.5, LLM-based chatbots' rationales highlighted the importance of scenario diversity, system familiarity, and efficient time management in testing.
\end{itemize}

%\begin{itemize}
%\item Regardless of their experience level, human testers consistently chose subsets of tests that covered dissimilar test scenarios. This reflects the widely accepted belief in the literature that a diverse test suite is effective at uncovering a variety of bugs.

%\item Two chatbots, Copilot and ChatGPT 4.0, made choices similar to human testers by prioritizing the execution of distinct test scenarios. This approach matched the rationale of most testers who preferred pairs of tests that were dissimilar.
    
%\item The other chatbots, ChatGPT 3.5 and Bard, selected similar test scenarios, focusing on ensuring coverage of the most frequent user interactions. Only a small fraction of testers (3.9\%) who chose similar test pairs cited the same rationale.
    
%\item LLM-based chatbots showed varied responses, with ChatGPT 3.5 in particular displaying variability when given the same prompt multiple times. However, their justifications consistently emphasized the value of covering a range of scenarios, gaining familiarity with the SUT during testing, and managing time constraints.
%\end{itemize}

Moving forward, an important question to address is, "How can LLMs reason about test diversity?" NLP techniques leverage the relationships between words and language, with semantics being a crucial aspect. Exploring this capability can open new avenues for LLMs to enhance Diversity-based Testing (DBT), allowing testers to focus on higher-level decision-making without needing to manage the intricacies of specific distance functions or parsing test artifacts.

To better understand the impact of interactions between Agentware and human testers on decision-making, the next step is to connect our findings to the SocraTest taxonomy by Feldt et al. \citep{feldt2023autonomous}. This will help us explore how Agentware's recommendations evolve over time to adapt to individual testers' ways of thinking.Our goals are to: (i) replicate this study with more models, variations and an interactive feedback loop between testers, (ii) investigate the autonomy levels of Agentware, and (iii) assess how ongoing collaboration influences the decision-making processes and intuition of both testers and chatbots over time. Those steps towards our vision aim to (i) \textit{explore} how the domain expertise evolves throughout those interactions cycles in a longitudinal exposure, and (ii) \textit{design} reference models to create, deploy and foster the adoption of Agentware for software testing.

\section*{Data Availability}

The complete data for this study, including (i) our survey, (ii) participants' answers, and (iii) the chatbots' answers, is available in a Zenodo package~\citep{replication2024aist}. The package includes instructions on the different files and scripts used in the study.

\bibliographystyle{ACM-Reference-Format}
\bibliography{bibliography}

\end{document}